\documentclass[conference]{IEEEtran}
\IEEEoverridecommandlockouts
\usepackage{cite}
\usepackage{amsmath,amssymb,amsfonts}
\usepackage{algorithmic}
\usepackage{graphicx}
\usepackage{textcomp}
\usepackage{xcolor}
\usepackage{multirow}
\def\BibTeX{{\rm B\kern-.05em{\sc i\kern-.025em b}\kern-.08em
    T\kern-.1667em\lower.7ex\hbox{E}\kern-.125emX}}
\begin{document}

\title{\fontsize{18}{20}\selectfont \textbf{CARMEN:} CORDIC-Accelerated Resource-Efficient\\ Multi-Precision Inference Engine for Deep Learning
}

\author{
\IEEEauthorblockN{Sonu Kumar, Mukul Lokhande, Santosh Kumar Vishvakarma}
 \IEEEauthorblockA{\textit{Dept. of Electrical Engineering} \\
 \textit{Indian Institute of Technology Indore} \\
 Simrol 453552, Indore, India \\
\{phd2101191002, phd2201102020, skvishvakarma\}@iiti.ac.in}\\
 \thanks{This work was supported partially by the Dept of Science and Technology (DST), Govt of India, for the INSPIRE PhD fellowship, and MeitY/SMDP-C2S for ASIC design tools.}

\and

 \IEEEauthorblockN{Adam Teman}
 \IEEEauthorblockA{\textit{EnICS Labs} \\
 \textit{Bar-Ilan University } \\
  Ramat Gan, Israel \\
 adam.teman@biu.ac.il}
 }

\maketitle

\begin{abstract}
This paper presents CARMEN, a runtime-adaptive, CORDIC-accelerated multi-precision vector engine for resource-efficient deep learning inference. The key insight is that CORDIC iteration depth directly governs computational accuracy, enabling dynamic switching between approximate and accurate execution modes without hardware modification. The architecture integrates a low-resource iterative CORDIC-based MAC unit with a time-multiplexed multi-activation function block, supporting flexible 8/16-bit precision and high hardware utilization. ASIC implementation in 28\,nm CMOS achieves up to 33\% reduction in computation cycles and 21\% power savings per MAC stage; a 256-PE configuration delivers 4.83 TOPS/mm\textsuperscript{2} compute density and 11.67 TOPS/W energy efficiency. FPGA deployment on Pynq-Z2 validates 154.6\,ms latency at 0.43\,W for real-time object detection.
\end{abstract}


\begin{table*}[!t]
\caption{SoTA Design approaches and comparison of respective design features in AI workloads}
\label{tab:sota-comp}
\renewcommand{\arraystretch}{1.35}
\resizebox{\textwidth}{!}{%
\begin{tabular}{|c|c|c|c|c|c|c|c|c|}
\hline
\textbf{Design} & \textbf{Baseline} & \textbf{ICIIS'25}~\cite{HYDRA_ICIIS25} & \textbf{ICIIS'25}~\cite{GR-Neuro} & \textbf{IEEE Access'24}~\cite{QuantMAC} & \textbf{TVLSI'25}~\cite{Flex-PE} & \textbf{ISCAS'25}~\cite{LPRE} & \textbf{ISVLSI'25}~\cite{NEURIC} & \textbf{Proposed} \\ \hline
\textbf{Compute} & Pipe-CORDIC & Pipe-CORDIC & \multicolumn{1}{c|}{PWL} & Pipe-CORDIC & Pipe-CORDIC & Logarithmic Approx. & Iterative CORDIC & Iterative CORDIC \\ \hline
\textbf{Arch. Type} & Fully Parallel & Layer-Reused & AF-Reused & AF-Reused & Systolic Array & \begin{tabular}[c]{@{}c@{}}Time-multiplexed \\ Reconfigurable Array\end{tabular} & Layer-Reused & Vector Engine \\ \hline
\textbf{Scalability} & No & Yes & No & No & - & Yes & No & Yes \\ \hline
\textbf{Precision} & FxP-8 & FxP-8 & FxP-8 & FxP-8 & FxP-4/8/16/32 & Posit-8/16/32 & FxP-8 & FxP-8/16 \\ \hline
\textbf{Accuracy loss} & High & High & High & High & Medium & Low & Medium & Variable (Low) \\ \hline
\textbf{Design Overhead} & Area, St. Power & Area & Area, St. Power & Area, Power & Energy & Area, Complexity & Latency & \begin{tabular}[c]{@{}c@{}}Application-optimized \\ Performance\end{tabular} \\ \hline
\textbf{AF-Supported} & ReLU & ReLU & Sigmoid/Tanh & NA & \begin{tabular}[c]{@{}c@{}}Sigmoid, Softmax, \\ Tanh, ReLU\end{tabular} & \begin{tabular}[c]{@{}c@{}}Sigmoid, Tanh, \\ Softmax\end{tabular} & Sigmoid/Tanh & \begin{tabular}[c]{@{}c@{}}SoftMax, GELU, Sigmoid\\ Tanh, Swish, ReLU, and SELU\end{tabular} \\ \hline
\textbf{Applications} & ANN & ANN & ANN & DNN & DNN, Transformers & DNN & DNN & DNN, Transformers (MLP) \\ \hline
\end{tabular}}
\end{table*}

\section{Introduction}
\IEEEPARstart{O}{ver} the past decade, deep learning has enabled transformative advancements across diverse application domains, driven by architectures such as Deep Neural Networks (DNNs), Vision Transformers (ViTs), and Large Language Models (LLMs)~\cite{Sze, RAMAN_N}. From early convolutional neural networks (CNNs) to large-scale transformer-based models, the fundamental computational kernels remain largely unchanged, consisting of convolutional, fully connected (FC) or multi-layer perceptron (MLP), and multi-head attention (MHA) layers~\cite{LSTM-AF-TC}. Workload characterization studies indicate that multiply-accumulate (MAC) operations dominate the computation, accounting for nearly 90\% of total operations, while activation functions (AFs) contribute only 2-5\%~\cite{TPUv4}. Efficient execution of these workloads under stringent area and power constraints remains a key challenge for edge AI deployment.

Prior works have explored several techniques to improve resource efficiency, including fixed-point quantization~\cite{QForce-RL}, CORDIC-based arithmetic~\cite{NEURIC}, logarithmic approximations~\cite{LPRE}, and truncation strategies~\cite{QuantMAC}. While these approaches reduce computational complexity\cite{XR-NPE}, they often introduce accuracy degradation of up to 4\%~\cite{QuantMAC} or lead to suboptimal energy efficiency~\cite{POLARON}. Furthermore, the imbalance in workload distribution between MAC operations and AFs has been largely overlooked. Existing designs allocate substantial hardware resources to AF units, resulting in poor utilization, with reports indicating up to 84\% idle cycles~\cite{HYDRA_ICIIS25} and 20-25\% area overhead due to inactive logic in accelerators such as Google TPUv4~\cite{TPUv4}. More fundamentally, the trade-off between computational accuracy and hardware efficiency is typically fixed at design time: static CORDIC iteration depths or approximation schemes either incur unnecessary latency for error-tolerant layers or compromise accuracy for sensitive computations. Table~\ref{tab:sota-comp} summarizes these limitations across current state-of-the-art (SoTA) approaches.

To address these gaps, this work adopts a runtime-adaptive design paradigm. The key insight is that CORDIC computation accuracy is directly governed by the number of iterations, which can be dynamically tuned without modifying the underlying hardware. This enables seamless switching between approximate and accurate execution modes with minimal area overhead. Additionally, by prioritizing MAC-dominated execution and time-multiplexing activation function hardware, the design significantly improves resource utilization.

Based on these principles, we present CARMEN, a CORDIC-accelerated multi-precision vector engine for resource-constrained deep learning applications. The key contributions are:
\begin{enumerate}
\item A runtime-adaptive iterative CORDIC-based MAC unit supporting approximate and accurate modes, enabling latency-accuracy trade-offs with up to 33\% cycle reduction and $<$2\% accuracy loss.

\item A reconfigurable multi-precision vector engine integrating MAC, time-multiplexed multi-AF (ReLU, GELU, SoftMax, Tanh, Sigmoid, Swish, SELU), pooling, and normalization units with efficient resource sharing and high hardware utilization.

\item Comprehensive FPGA and ASIC evaluation in 28\,nm CMOS, demonstrating 4.83\,TOPS/mm\textsuperscript{2} compute density and 11.67\,TOPS/W energy efficiency, with system-level validation at 154.6\,ms latency and 0.43\,W on Pynq-Z2.
\end{enumerate}

\section{Proposed Solution}

\begin{figure}[t]
    \centering
   \includegraphics[width=\columnwidth]{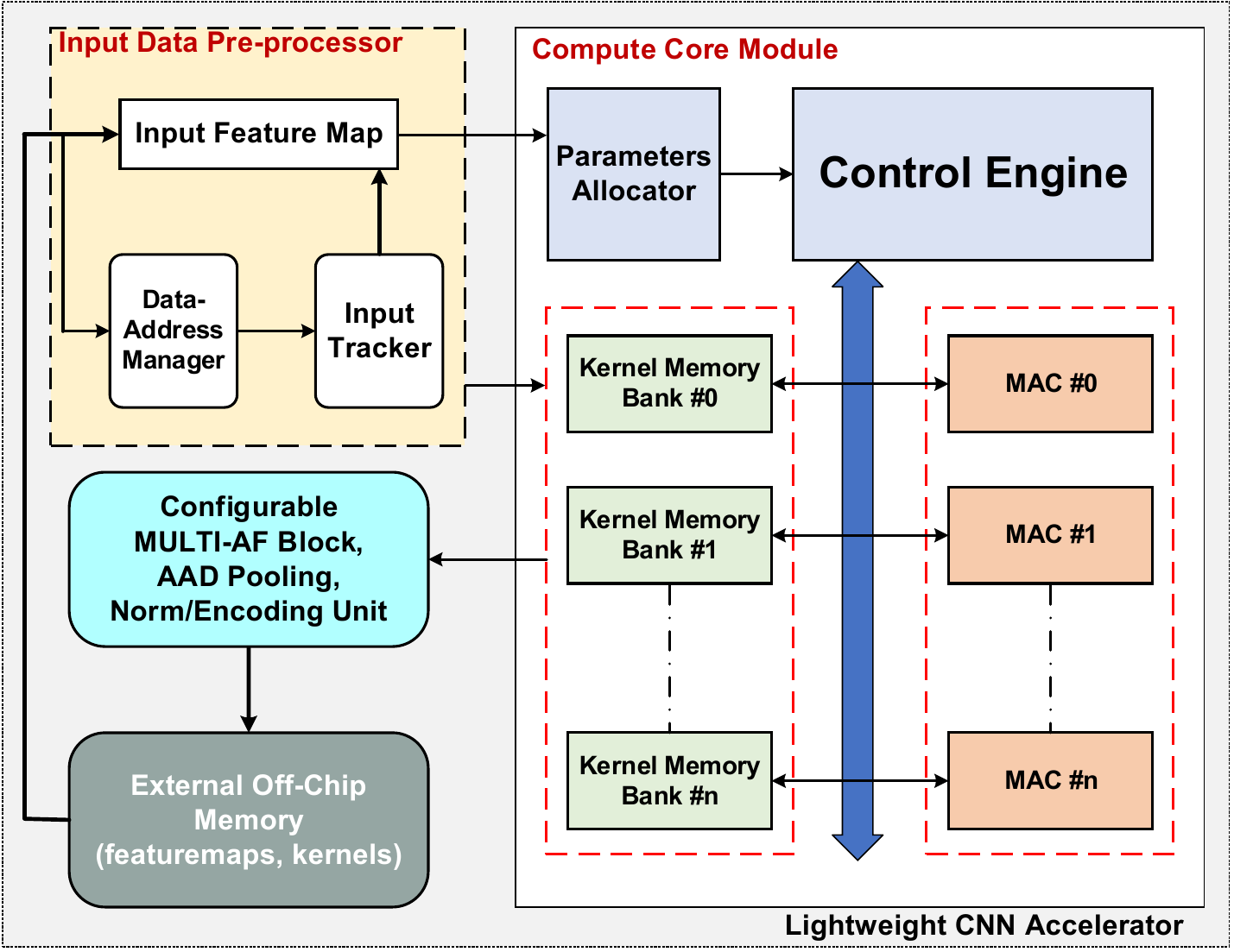}
    \caption{Resource-efficient DL accelerator, comprised of Vector Engine, data pre-fetcher, control engine, interface and off-chip memory.}
    \label{fig:arch}
\end{figure}

The architecture of the proposed CARMEN accelerator is shown in Fig.~\ref{fig:arch}. It consists of five key components: (i)~a runtime-configurable vector engine~\cite{HYDRA_ICIIS25}, (ii)~a time-multiplexed multi-AF unit~\cite{NEURIC} supporting Tanh, Softmax, GELU, Sigmoid, Swish, ReLU, and SELU, (iii)~a performance-enhanced AAD pooling unit~\cite{AAD-pool}, (iv)~a normalization unit, and (v)~a data pre-processing and off-chip memory interface. The vector engine forms the compute core and comprises $N$ precision-adjustable, accuracy-configurable iterative MAC units, each supplied by dual kernel memory banks of size ($n$-bit $\times$ 32) for weights and input activations. A hierarchical, synthesizable generator framework enables scalable instantiation of the architecture for both edge and high-performance computing (HPC) scenarios.

\subsection{Runtime-Adaptive Low-Resource Iterative CORDIC MAC}

The unified CORDIC algorithm, introduced by Walther~\cite{Walther1971}, enables efficient computation of circular, linear, and hyperbolic functions using simple arithmetic components such as adders/subtractors, multiplexers, and shift units. Recent works, including ReCON~\cite{RECON} and Flex-PE~\cite{Flex-PE}, have leveraged CORDIC for DNN primitives such as MAC operations and nonlinear activation functions. However, such unified CORDIC-based vector engines are inherently general-purpose and do not exploit the workload imbalance in DNNs, where AFs contribute only 2-5\% of operations. As a result, hyperbolic and circular rotation units dedicated to AF computation remain largely idle, incurring unnecessary area and power overhead.

\begin{figure}[!t]
    \centering
    \includegraphics[width=\columnwidth]{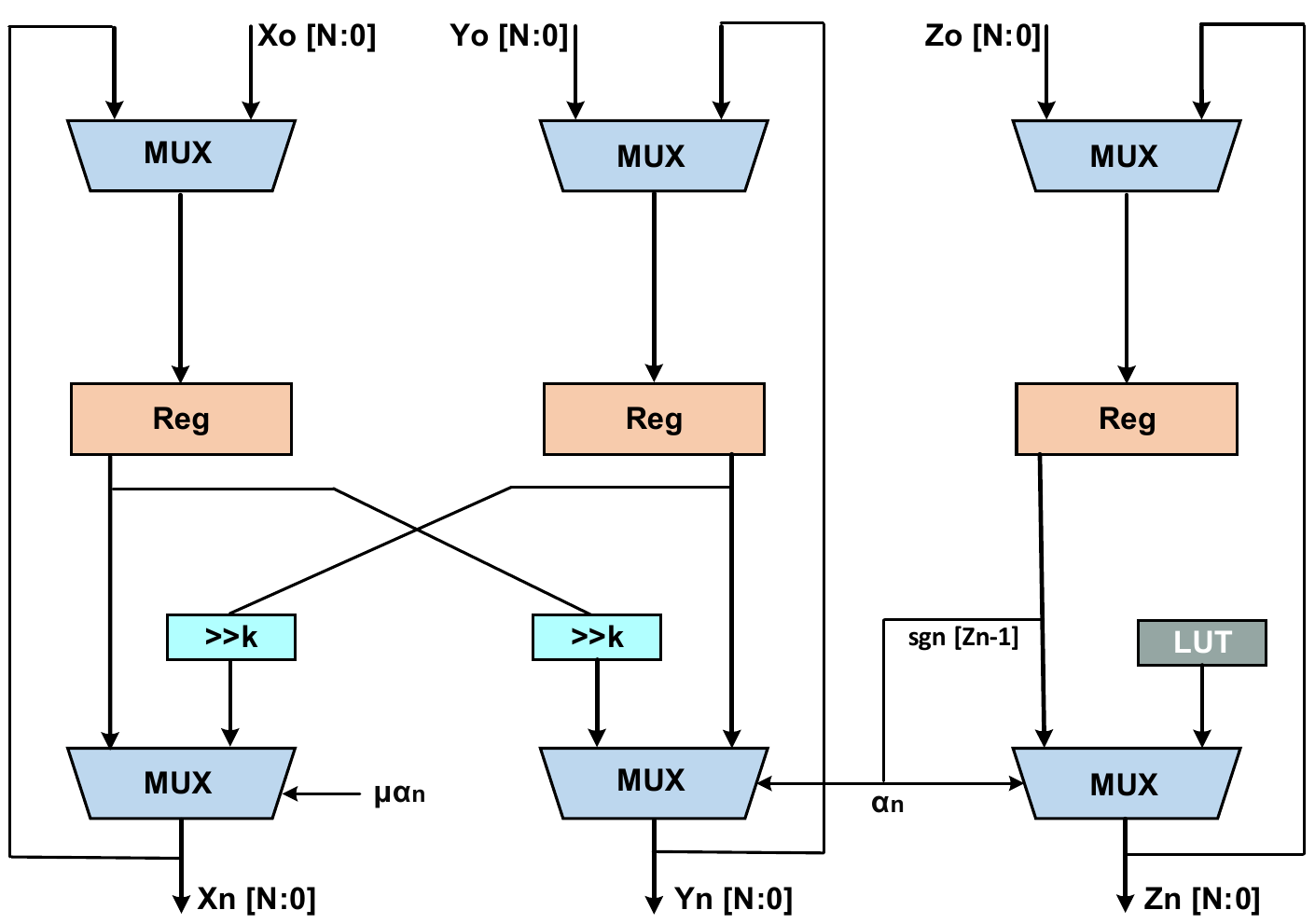}
    \caption{Iterative low-latency CORDIC MAC architecture.}
    \label{fig:iter-cordic-mac}
\end{figure}

To overcome this inefficiency, we propose an optimized iterative CORDIC-based MAC architecture, as illustrated in Fig.~\ref{fig:iter-cordic-mac}. Unlike conventional pipelined implementations that replicate hardware across stages, the proposed design reuses arithmetic resources across iterations, thereby significantly reducing area and switching activity. This approach achieves up to 33\% reduction in computation cycles and 21\% lower power per stage. Although iterative execution introduces additional latency per MAC operation, this overhead is effectively amortized through parallel vector execution, making the design well-suited for high-throughput accelerator systems.

\subsection{Dynamically Configurable Versatile Multi-AF Block}

The proposed multi-AF unit improves hardware efficiency through time-multiplexed execution of multiple nonlinear functions. Instead of dedicating separate hardware resources for each activation function, the design reuses shared CORDIC datapaths along with lightweight auxiliary logic. The CORDIC core operates in hyperbolic rotation mode to compute sigmoid and tanh, while softmax is realized through a combination of exponentiation and normalization stages. Simpler functions such as ReLU and its variants are handled by bypass logic with negligible overhead. Additional components such as multiplexers, FIFO buffers, and small multipliers enable support for the full set of activation functions with less than 4\% area overhead. Since AFs account for only 2-5\% of total operations, the time-multiplexed schedule does not bottleneck overall throughput.

\subsection{Peripherals: AAD Pooling and Normalization Unit}

The architecture includes supporting peripherals: an on-the-fly AAD pooling unit~\cite{AAD-pool}, a normalization unit, and an AXI-based off-chip memory interface. AAD pooling improves accuracy by 0.5-1\% in CORDIC-based designs~\cite{Flex-PE} while maintaining lower computational complexity and hardware overhead compared to conventional pooling techniques~\cite{AAD-pool}.

The control engine, inspired by prior work~\cite{Flex-PE}, includes configuration registers for runtime parameter tuning, status and control flags for coordination with the FSMD control path, and an address mapping unit for efficient data movement. This enables flexible execution across different layers and workloads while maintaining high hardware utilization.

\begin{table*}[!t]
\caption{Comparative Performance Metrics for Different CORDIC-based SoTA MAC Units}
\label{tab:mac-util-comp}
\renewcommand{\arraystretch}{1.15}
\resizebox{\textwidth}{!}{%
\begin{tabular}{|cccccccccccccc|}
\hline
\multicolumn{1}{|c|}{\textbf{Design}} & \multicolumn{2}{c|}{\textbf{TCAS-II'24~\cite{RPE}}} & \multicolumn{2}{c|}{\textbf{ISCAS'25~\cite{LPRE}}} & \multicolumn{5}{c|}{\textbf{ICIIS'25~\cite{HYDRA_ICIIS25}}} & \multicolumn{1}{c|}{\textbf{TVLSI'25~\cite{MSDF-MAC}}} & \multicolumn{1}{c|}{\textbf{TCAD'22~\cite{Acc-App-MAC}}} & \multicolumn{1}{c|}{\textbf{TVLSI'25~\cite{Flex-PE}}} & \textbf{Proposed} \\ \hline
\multicolumn{14}{|c|}{{\textbf{FPGA Utilization (VC707, 100 MHz)}}} \\ \hline
\multicolumn{1}{|c|}{\textbf{Parameter}} & \multicolumn{1}{c|}{\textbf{FP32}} & \multicolumn{1}{c|}{\textbf{FP32}} & \multicolumn{1}{c|}{\textbf{BF16}} & \multicolumn{1}{c|}{\textbf{Posit-8}} & \multicolumn{1}{c|}{\textbf{Vedic}} & \multicolumn{1}{c|}{\textbf{Wallace}} & \multicolumn{1}{c|}{\textbf{Booth}} & \multicolumn{1}{c|}{\textbf{Quant-MAC}} & \multicolumn{1}{c|}{\textbf{CORDIC}} & \multicolumn{1}{c|}{\textbf{MSDF-MAC}} & \multicolumn{1}{c|}{\textbf{Acc-App-MAC}} & \multicolumn{1}{c|}{\textbf{CORDIC}} & \textbf{Iter-MAC} \\ \hline
\multicolumn{1}{|c|}{\textbf{LUTs}} & \multicolumn{1}{c|}{8065} & \multicolumn{1}{c|}{8054} & \multicolumn{1}{c|}{3670} & \multicolumn{1}{c|}{467} & \multicolumn{1}{c|}{160} & \multicolumn{1}{c|}{106} & \multicolumn{1}{c|}{84} & \multicolumn{1}{c|}{72} & \multicolumn{1}{c|}{56} & \multicolumn{1}{c|}{62} & \multicolumn{1}{c|}{57} & \multicolumn{1}{c|}{45} & 24 \\ \hline
\multicolumn{1}{|c|}{\textbf{FFs}} & \multicolumn{1}{c|}{1072} & \multicolumn{1}{c|}{1718} & \multicolumn{1}{c|}{324} & \multicolumn{1}{c|}{175} & \multicolumn{1}{c|}{241} & \multicolumn{1}{c|}{113} & \multicolumn{1}{c|}{59} & \multicolumn{1}{c|}{56} & \multicolumn{1}{c|}{72} & \multicolumn{1}{c|}{45} & \multicolumn{1}{c|}{NR} & \multicolumn{1}{c|}{37} & 22 \\ \hline
\multicolumn{1}{|c|}{\textbf{Delay (ns)}} & \multicolumn{1}{c|}{5.56} & \multicolumn{1}{c|}{4.6} & \multicolumn{1}{c|}{0.512} & \multicolumn{1}{c|}{2.68} & \multicolumn{1}{c|}{4.5} & \multicolumn{1}{c|}{2.6} & \multicolumn{1}{c|}{3.1} & \multicolumn{1}{c|}{5.4} & \multicolumn{1}{c|}{1.52} & \multicolumn{1}{c|}{3.2} & \multicolumn{1}{c|}{3.51} & \multicolumn{1}{c|}{4.5} & 9.1 \\ \hline
\multicolumn{1}{|c|}{\textbf{Power (mW)}} & \multicolumn{1}{c|}{378} & \multicolumn{1}{c|}{296} & \multicolumn{1}{c|}{136} & \multicolumn{1}{c|}{68} & \multicolumn{1}{c|}{6.1} & \multicolumn{1}{c|}{3.3} & \multicolumn{1}{c|}{3.1} & \multicolumn{1}{c|}{4.2} & \multicolumn{1}{c|}{8.3} & \multicolumn{1}{c|}{5.8} & \multicolumn{1}{c|}{6.9} & \multicolumn{1}{c|}{2} & 1.9 \\ \hline
\multicolumn{1}{|c|}{\textbf{PDP (pJ)}} & \multicolumn{1}{c|}{2102} & \multicolumn{1}{c|}{1361.6} & \multicolumn{1}{c|}{69.6} & \multicolumn{1}{c|}{182} & \multicolumn{1}{c|}{27.45} & \multicolumn{1}{c|}{8.58} & \multicolumn{1}{c|}{9.6} & \multicolumn{1}{c|}{22.68} & \multicolumn{1}{c|}{12.6} & \multicolumn{1}{c|}{18.56} & \multicolumn{1}{c|}{24.2} & \multicolumn{1}{c|}{9} & 17.29 \\ \hline
\multicolumn{14}{|c|}{{\textbf{ASIC Performance (28nm, 0.9V)}}} \\ \hline
\multicolumn{1}{|c|}{\textbf{Area ($\mu$m$^2$)}} & \multicolumn{1}{c|}{10000} & \multicolumn{1}{c|}{13000} & \multicolumn{1}{c|}{4340} & \multicolumn{1}{c|}{754} & \multicolumn{1}{c|}{407} & \multicolumn{1}{c|}{296} & \multicolumn{1}{c|}{271} & \multicolumn{1}{c|}{175} & \multicolumn{1}{c|}{264} & \multicolumn{1}{c|}{286} & \multicolumn{1}{c|}{259} & \multicolumn{1}{c|}{8570} & 108 \\ \hline
\multicolumn{1}{|c|}{\textbf{Delay (ns)}} & \multicolumn{1}{c|}{679} & \multicolumn{1}{c|}{700} & \multicolumn{1}{c|}{295} & \multicolumn{1}{c|}{40.6} & \multicolumn{1}{c|}{6.38} & \multicolumn{1}{c|}{5.62} & \multicolumn{1}{c|}{5.3} & \multicolumn{1}{c|}{3.58} & \multicolumn{1}{c|}{2.36} & \multicolumn{1}{c|}{1.42} & \multicolumn{1}{c|}{2.6} & \multicolumn{1}{c|}{0.7} & 2.98 \\ \hline
\multicolumn{1}{|c|}{\textbf{Power (mW)}} & \multicolumn{1}{c|}{15.86} & \multicolumn{1}{c|}{29.3} & \multicolumn{1}{c|}{6.89} & \multicolumn{1}{c|}{1.8} & \multicolumn{1}{c|}{35} & \multicolumn{1}{c|}{37} & \multicolumn{1}{c|}{12.8} & \multicolumn{1}{c|}{89} & \multicolumn{1}{c|}{24.5} & \multicolumn{1}{c|}{6.7} & \multicolumn{1}{c|}{12.4} & \multicolumn{1}{c|}{1.5} & 6.3 \\ \hline
\multicolumn{1}{|c|}{\textbf{PDP (pJ)}} & \multicolumn{1}{c|}{10768.94} & \multicolumn{1}{c|}{20510} & \multicolumn{1}{c|}{4682} & \multicolumn{1}{c|}{1189} & \multicolumn{1}{c|}{223.3} & \multicolumn{1}{c|}{207.94} & \multicolumn{1}{c|}{67.84} & \multicolumn{1}{c|}{318.62} & \multicolumn{1}{c|}{57.82} & \multicolumn{1}{c|}{9.514} & \multicolumn{1}{c|}{32.24} & \multicolumn{1}{c|}{1.05} & 18.774 \\ \hline
\end{tabular}}
\end{table*}

\begin{table*}[!t]
\caption{Comparative Performance Metrics for Different CORDIC-based SoTA AF Units}
\label{tab:resource-af}
\renewcommand{\arraystretch}{1.45}
\resizebox{\textwidth}{!}{%
\begin{tabular}{|ccccccccccccccc|}
\hline
\multicolumn{1}{|c|}{\textbf{Design}} & \multicolumn{3}{c|}{\textbf{ISQED'24~\cite{FP-CORDIC-ISQED24}}} & \multicolumn{1}{c|}{\textbf{TCAS-II'20~\cite{Precision-Scalable-AF-TCAS-II'20}}} & \multicolumn{1}{c|}{\textbf{TVLSI'23~\cite{AxAF-TVLSI'23}}} & \multicolumn{3}{c|}{\textbf{ISQED'24~\cite{FP-CORDIC-ISQED24}}} & \multicolumn{1}{c|}{\textbf{TC'23~\cite{LSTM-AF-TC}}} & \multicolumn{3}{c|}{\textbf{ISQED'24~\cite{FP-CORDIC-ISQED24}}} & \multicolumn{1}{c|}{\textbf{TVLSI'25~\cite{Flex-PE}}} & \textbf{Proposed} \\ \hline
\multicolumn{15}{|c|}{{\textbf{FPGA Utilization (VC707, 100 MHz)}}} \\ \hline
\multicolumn{1}{|c|}{\textbf{Parameter}} & \multicolumn{1}{c|}{Softmax-FP32} & \multicolumn{1}{c|}{Softmax-FP16} & \multicolumn{1}{c|}{Softmax-BF16} & \multicolumn{1}{c|}{Softmax- FxP8/16} & \multicolumn{1}{c|}{Softmax-16b} & \multicolumn{1}{c|}{Tanh-FP32} & \multicolumn{1}{c|}{Tanh-FP16} & \multicolumn{1}{c|}{Tanh-BF16} & \multicolumn{1}{c|}{Tanh/Sigmoid-16b} & \multicolumn{1}{c|}{Sigmoid-FP32} & \multicolumn{1}{c|}{Sigmoid-FP16} & \multicolumn{1}{c|}{Sigmoid-BF16} & \multicolumn{1}{c|}{SSTp} & FxP8/16 \\ \hline
\multicolumn{1}{|c|}{\textbf{LUTs}} & \multicolumn{1}{c|}{3217} & \multicolumn{1}{c|}{1137} & \multicolumn{1}{c|}{1263} & \multicolumn{1}{c|}{2564} & \multicolumn{1}{c|}{1215} & \multicolumn{1}{c|}{4298} & \multicolumn{1}{c|}{1530} & \multicolumn{1}{c|}{1513} & \multicolumn{1}{c|}{2395} & \multicolumn{1}{c|}{5101} & \multicolumn{1}{c|}{1853} & \multicolumn{1}{c|}{1856} & \multicolumn{1}{c|}{897} & 537 \\ \hline
\multicolumn{1}{|c|}{\textbf{FFs}} & \multicolumn{1}{c|}{NR} & \multicolumn{1}{c|}{NR} & \multicolumn{1}{c|}{NR} & \multicolumn{1}{c|}{2794} & \multicolumn{1}{c|}{1012} & \multicolumn{1}{c|}{NR} & \multicolumn{1}{c|}{NR} & \multicolumn{1}{c|}{NR} & \multicolumn{1}{c|}{1503} & \multicolumn{1}{c|}{NR} & \multicolumn{1}{c|}{NR} & \multicolumn{1}{c|}{NR} & \multicolumn{1}{c|}{1231} & 468 \\ \hline
\multicolumn{1}{|c|}{\textbf{Delay (ns)}} & \multicolumn{1}{c|}{92} & \multicolumn{1}{c|}{43} & \multicolumn{1}{c|}{45} & \multicolumn{1}{c|}{2.3} & \multicolumn{1}{c|}{3.32} & \multicolumn{1}{c|}{56} & \multicolumn{1}{c|}{34} & \multicolumn{1}{c|}{38} & \multicolumn{1}{c|}{0.18} & \multicolumn{1}{c|}{109} & \multicolumn{1}{c|}{60} & \multicolumn{1}{c|}{45} & \multicolumn{1}{c|}{11.8} & 2.6 \\ \hline
\multicolumn{1}{|c|}{\textbf{Power (mW)}} & \multicolumn{1}{c|}{115} & \multicolumn{1}{c|}{115} & \multicolumn{1}{c|}{77} & \multicolumn{1}{c|}{NR} & \multicolumn{1}{c|}{165} & \multicolumn{1}{c|}{130} & \multicolumn{1}{c|}{124} & \multicolumn{1}{c|}{82} & \multicolumn{1}{c|}{681} & \multicolumn{1}{c|}{121} & \multicolumn{1}{c|}{118} & \multicolumn{1}{c|}{83} & \multicolumn{1}{c|}{59} & 30 \\ \hline
\multicolumn{1}{|c|}{\textbf{PDP (pJ)}} & \multicolumn{1}{c|}{10580} & \multicolumn{1}{c|}{4945} & \multicolumn{1}{c|}{3465} & \multicolumn{1}{c|}{-} & \multicolumn{1}{c|}{548} & \multicolumn{1}{c|}{7280} & \multicolumn{1}{c|}{4216} & \multicolumn{1}{c|}{3116} & \multicolumn{1}{c|}{123} & \multicolumn{1}{c|}{13189} & \multicolumn{1}{c|}{7080} & \multicolumn{1}{c|}{3735} & \multicolumn{1}{c|}{696.2} & 78 \\ \hline
\multicolumn{15}{|c|}{{ \textbf{ASIC Performance (28nm, 0.9V)}}} \\ \hline
\multicolumn{1}{|c|}{\textbf{Area ($\mu$m$^2$)}} & \multicolumn{1}{c|}{41536} & \multicolumn{1}{c|}{17289} & \multicolumn{1}{c|}{11301} & \multicolumn{1}{c|}{18392} & \multicolumn{1}{c|}{3819} & \multicolumn{1}{c|}{5060} & \multicolumn{1}{c|}{1180} & \multicolumn{1}{c|}{843} & \multicolumn{1}{c|}{870523} & \multicolumn{1}{c|}{2234} & \multicolumn{1}{c|}{1855} & \multicolumn{1}{c|}{1180} & \multicolumn{1}{c|}{49152} & 2138 \\ \hline
\multicolumn{1}{|c|}{\textbf{Delay (ns)}} & \multicolumn{1}{c|}{6} & \multicolumn{1}{c|}{4} & \multicolumn{1}{c|}{3.3} & \multicolumn{1}{c|}{0.3} & \multicolumn{1}{c|}{1.6} & \multicolumn{1}{c|}{4} & \multicolumn{1}{c|}{3.3} & \multicolumn{1}{c|}{3.4} & \multicolumn{1}{c|}{NR} & \multicolumn{1}{c|}{7.6} & \multicolumn{1}{c|}{4.4} & \multicolumn{1}{c|}{3.26} & \multicolumn{1}{c|}{2.3} & 2.6 \\ \hline
\multicolumn{1}{|c|}{\textbf{Power (mW)}} & \multicolumn{1}{c|}{75} & \multicolumn{1}{c|}{40} & \multicolumn{1}{c|}{25} & \multicolumn{1}{c|}{51.6} & \multicolumn{1}{c|}{1.6} & \multicolumn{1}{c|}{8.75} & \multicolumn{1}{c|}{3} & \multicolumn{1}{c|}{2} & \multicolumn{1}{c|}{150} & \multicolumn{1}{c|}{10} & \multicolumn{1}{c|}{4.8} & \multicolumn{1}{c|}{2.5} & \multicolumn{1}{c|}{5.2} & 60 \\ \hline
\multicolumn{1}{|c|}{\textbf{PDP (pJ)}} & \multicolumn{1}{c|}{450} & \multicolumn{1}{c|}{160} & \multicolumn{1}{c|}{82.5} & \multicolumn{1}{c|}{15.5} & \multicolumn{1}{c|}{2.56} & \multicolumn{1}{c|}{35} & \multicolumn{1}{c|}{9.9} & \multicolumn{1}{c|}{6.8} & \multicolumn{1}{c|}{-} & \multicolumn{1}{c|}{76} & \multicolumn{1}{c|}{21.12} & \multicolumn{1}{c|}{8.15} & \multicolumn{1}{c|}{11.96} & 156 \\ \hline
\end{tabular}}
\end{table*}

\section{Methodology \& Performance Evaluation}

\begin{figure}[!t]
    \centering
    \includegraphics[width=\columnwidth]{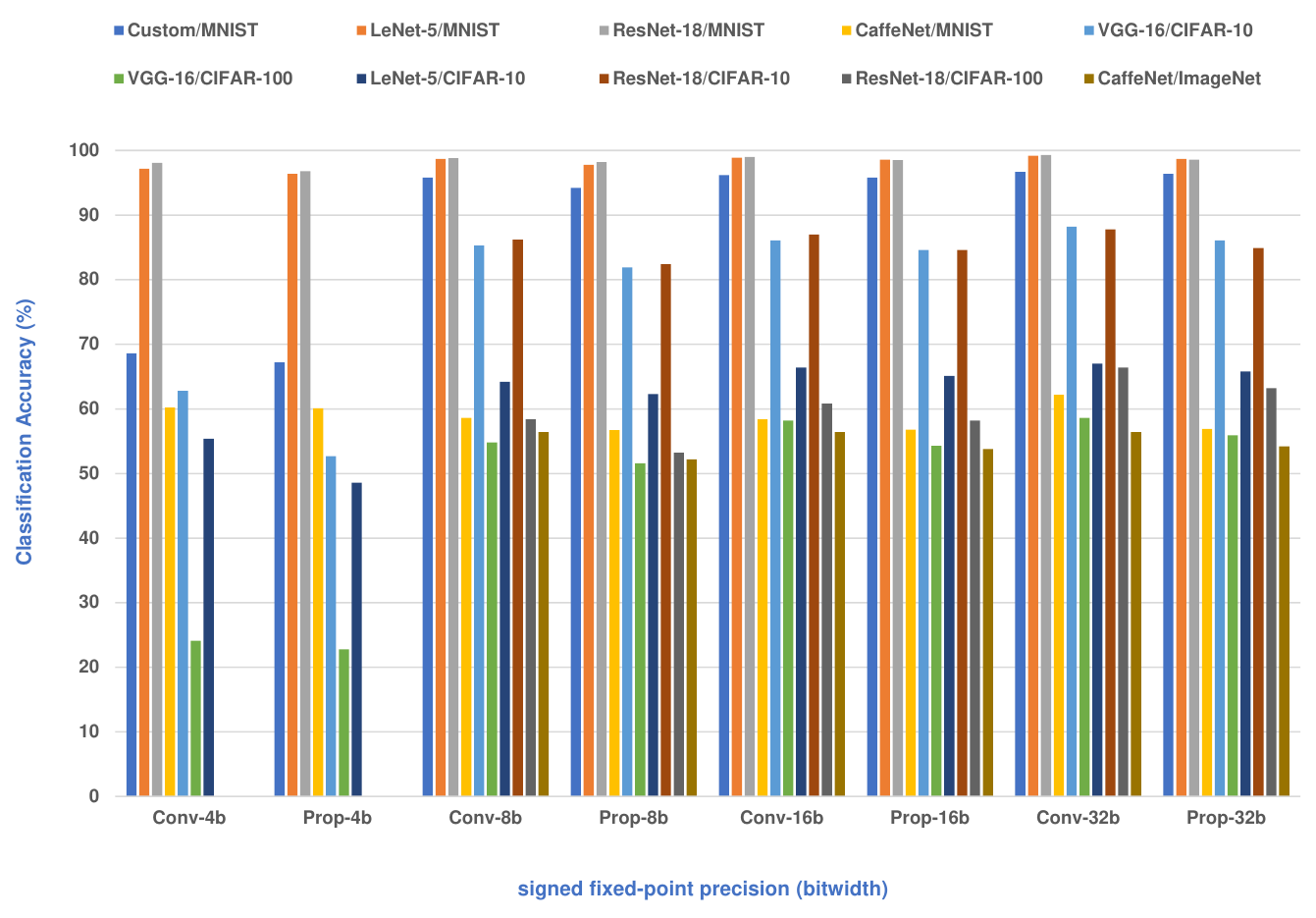}
    \caption{DNN accuracy evaluation across representative models using CORDIC-based computation.}
    \label{fig:accuracy}
\end{figure}

The proposed vector engine is evaluated using a hardware-software co-design framework that integrates algorithm-level validation with hardware implementation analysis. A cycle-accurate software model is developed in Python using FxP-Math and QKeras to emulate custom CORDIC-based arithmetic and configurable neural network layers. This framework enables systematic evaluation across multiple precision formats, layer configurations, and workload types, with direct comparison against an FP32 baseline. Fig.~\ref{fig:accuracy} shows the impact of precision scaling and CORDIC iteration depth on inference accuracy across representative DNN models. At 8-bit fixed-point precision, the CORDIC-based computation maintains accuracy within 2\% of the conventional baseline for all evaluated networks, with negligible degradation at 16-bit and above.

To validate architectural flexibility, the parameterized, precision-adjustable, and stage-configurable iterative MAC unit is implemented in Verilog HDL. The number of processing elements (PEs) per layer is defined at design time, allowing scalable generation of the vector engine, while control and sequencing are managed by a centralized control engine. The outputs from the hardware-aware Python model are cross-validated against RTL simulations using the Synopsys VCS simulator to ensure functional correctness. Iterative compute stages for each layer are configured using an accuracy-sensitivity metric~\cite{Flex-PE}, enabling dynamic selection between approximate and accurate modes based on layer criticality.

FPGA synthesis and implementation are performed using the AMD Vivado Design Suite targeting the VC707 platform. Resource utilization results for the MAC units and multi-AF block are reported in Table~\ref{tab:mac-util-comp} and Table~\ref{tab:resource-af}, respectively. In addition, the designs are synthesized using Synopsys Design Compiler in a 28\,nm CMOS HPC+ technology to obtain ASIC-level performance metrics.

Table~\ref{tab:mac-util-comp} shows the hardware efficiency of the iterative MAC unit. By reusing arithmetic resources across iterations, CARMEN requires only 24 LUTs and 22 FFs on FPGA roughly half the resources of the closest CORDIC-based competitor~\cite{Flex-PE} (45 LUTs, 37 FFs), while consuming 1.9\,mW. In the 28\,nm ASIC implementation, the iterative MAC occupies just 108\,$\mu$m$^2$, the smallest among all compared designs. Although iterative execution increases per-operation latency, this is amortized through parallel execution across multiple PEs, and the reduced switching activity yields lower dynamic power.

Table~\ref{tab:resource-af} compares activation function implementations across SoTA architectures. The time-multiplexed multi-AF block requires 537 LUTs and 468 FFs to support seven activation functions, compared to 897 LUTs and 1231 FFs for the single-function SSTp block in~\cite{Flex-PE}. On ASIC, the multi-AF unit occupies 2138\,$\mu$m$^2$, which is over 20$\times$ smaller than the CORDIC-based LSTM AF in~\cite{LSTM-AF-TC}, while achieving a PDP of 78\,pJ on FPGA. The resource savings stem from sharing CORDIC datapaths across functions rather than replicating hardware for each, with only minor control overhead.

\section{Vector Engine Generation \& System Evaluation}

The vector engine is evaluated at the system level using both FPGA and ASIC implementations under realistic workload conditions. Performance comparisons with prior works are summarized in Tables~\ref{tab:arch-fpga} and~\ref{tab:arch-asic}. Two configurations are considered to demonstrate scalability: (i)~a 64-PE design for compute-equivalent comparison, and (ii)~a 256-PE design for resource-equivalent comparison.

To validate robustness, prior state-of-the-art compute units, including LPRE~\cite{LPRE}, ILM~\cite{ILM}, and PWL~\cite{GR-Neuro}, are integrated into the same vector engine framework, enabling unified evaluation across heterogeneous compute paradigms. The CORDIC-based iterative MAC consistently outperforms these alternatives in both resource efficiency and inference accuracy, surpassing FP- and Posit-based implementations while maintaining high computational throughput.

FPGA results (Table~\ref{tab:arch-fpga}) show that CARMEN achieves 6.43\,GOPS/W energy efficiency at only 0.53\,W and 26.7k LUTs with zero DSP usage. Compared to Flex-PE~\cite{Flex-PE}, which achieves higher energy efficiency (8.42\,GOPS/W), it does so at 4.2$\times$ higher power (2.24\,W) and requires 73 DSP blocks. Among DSP-free designs, CARMEN outperforms LPRE~\cite{LPRE} (2.64\,GOPS/W, 1.6\,W) by 2.4$\times$ in energy efficiency. The lower operating frequency (85.4\,MHz) compared to systolic architectures is offset by reduced area and power overhead, underscoring the advantage of resource-efficient design for edge AI.

\begin{table*}[!t]
\caption{Analysis of FPGA Hardware Implementation for object detection (TinyYolo-v3) with SoTA AI accelerator designs}
\label{tab:arch-fpga}
\renewcommand{\arraystretch}{1.05}
\resizebox{\textwidth}{!}{%
\begin{tabular}{|c|c|c|c|c|c|c|c|c|}
\hline
\textbf{Design} & \textbf{Platform} & \textbf{Precision} & \textbf{k-LUTs} & \textbf{k-Regs/FFs} & \textbf{DSPs} & \textbf{\begin{tabular}[c]{@{}c@{}}Op. Freq \\      (MHz)\end{tabular}} & \textbf{\begin{tabular}[c]{@{}c@{}}Energy efficiency \\      (GOPS/W)\end{tabular}} & \textbf{Power(W)} \\ \hline
\textbf{Proposed} & VC707 & 8/16 & 26.7 & 15.9 & - & 85.4 & 6.43 & 0.53 \\ \hline
\textbf{TVLSI'25~\cite{Flex-PE}} & VC707 & 4/8/16/32 & 38.7 & 17.4 & 73 & 466 & 8.42 & 2.24 \\ \hline
\textbf{TCAS-I'24~\cite{BWu-TCASI24}} & ZU3EG & 8 & 40.8 & 45.5 & 258 & 100 & 0.39 & 2.2 \\ \hline
\textbf{TCAS-II'23~\cite{Ski-TCASII'23}} & XCVU9P & 8 & 132 & 39.5 & 96 & 150 & 6.36 & 5.52 \\ \hline
\textbf{TVLSI'23~\cite{lee-TVLSI}} & ZCU102 & 8 & 117 & 74 & 132 & 300 & 4.2 & 6.58 \\ \hline
\textbf{Access'24~\cite{QuantMAC}} & VC707 & 4/8 & 19.8 & 12.1 & 39 & 136 & 0.68 & 1.81 \\ \hline
\textbf{ISCAS'25~\cite{LPRE}} & VCU129 & 8/16/32 & 17.5 & 14.8 & - & 54.5 & 2.64 & 1.6 \\ \hline
\end{tabular}}
\end{table*}

\begin{table*}[!t]
\caption{ASIC Performance Comparison with SoTA 8-bit accelerator designs, with CMOS 28nm, 0.9V, SF technology.}
\label{tab:arch-asic}
\renewcommand{\arraystretch}{1.2}
\resizebox{\textwidth}{!}{%
\begin{tabular}{|l|c|l|c|c|c|c|c|}
\hline
\textbf{Design} & \textbf{Network/Arch} & \textbf{Datatype} & \textbf{Freq. (GHz)} & \textbf{Area (mm\textsuperscript{2})} & \textbf{Power (mW)} & \begin{tabular}[c]{@{}c@{}}\textbf{Energy Efficiency}\\ \textbf{TOPS/W}\end{tabular} & \begin{tabular}[c]{@{}c@{}}\textbf{Compute Density}\\ TOPS/mm\textsuperscript{2}\end{tabular} \\ \hline

\multirow{2}{*}{\textbf{TCAS-II'24~\cite{RPE}}} & \multirow{2}{*}{Vector Engine (64$\times$MACs)} & \multirow{2}{*}{FP8} & 1.47 & 0.896 & 1622 & 7.24 & 2.39 \\ \cline{4-8} 
 &  &  & 1.29 & 1.18 & 1375 & 3.57 & 1.21 \\ \hline
 \textbf{TCAS-I'22~\cite{ILM}} & \begin{tabular}[c]{@{}c@{}}Vector Engine (64$\times$MACs)\\ 196-64-32-32-10\end{tabular} & INT-8 & 0.4 & 2.43 & 224.6 & 7.75 & 1.67 \\ \hline
\textbf{ISCAS'25~\cite{LPRE}} & \begin{tabular}[c]{@{}c@{}}TREA (64$\times$MACs)\\ 196-64-32-32-10\end{tabular} & Posit-8 & 1.25 & 6.73 & 230.4 & 7.55 & 0.16 \\ \hline
 \textbf{TVLSI'25~\cite{Flex-PE}} & Systolic Array (8x8) & FxP8& 0.44 & 1.85 & 523 & 4.3 & 2.76 \\ \hline
\textbf{ICIIS'25~\cite{HYDRA_ICIIS25}} & \begin{tabular}[c]{@{}c@{}}Layer-Reused (64$\times$MACs)\\ 196-64-32-32-10\end{tabular} & FxP8 & 0.25 & 3.78 & 1540 & 4.28 & 2.07 \\ \hline
\multirow{2}{*}{\textbf{Proposed}} & Vector Engine 64$\times$PEs & \multirow{2}{*}{FxP8} & 1.24 & 0.43 & 329 & 3.84 & 1.52 \\ \cline{2-2} \cline{4-8} 
 & Vector Engine 256$\times$PEs &  & 0.96 & 1.42 & 1186 & 11.67 & 4.83 \\ \hline
 \textbf{Access'24~\cite{QuantMAC}} & \begin{tabular}[c]{@{}c@{}}Shared Bank (256$\times$MACs)\\ 784-196-120-84-10\end{tabular} & FxP8 & 0.28 & 1.58 & 499.7 & 6.87 & 1.18 \\ \hline
\end{tabular}}
\end{table*}

ASIC results (Table~\ref{tab:arch-asic}) further highlight the benefits of the iterative MAC design. The 64-PE configuration occupies only 0.43\,mm$^2$  the smallest footprint among all compared accelerators, while achieving 3.84\,TOPS/W. Scaling to 256 PEs yields 4.83\,TOPS/mm$^2$ compute density and 11.67\,TOPS/W energy efficiency, representing 1.75$\times$ and 1.5$\times$ improvements over the closest competitors~\cite{Flex-PE, RPE}, respectively. This near-linear scaling with comparable power consumption validates the architectural scalability.

\begin{figure}[!t]
    \centering
    \includegraphics[width=\columnwidth]{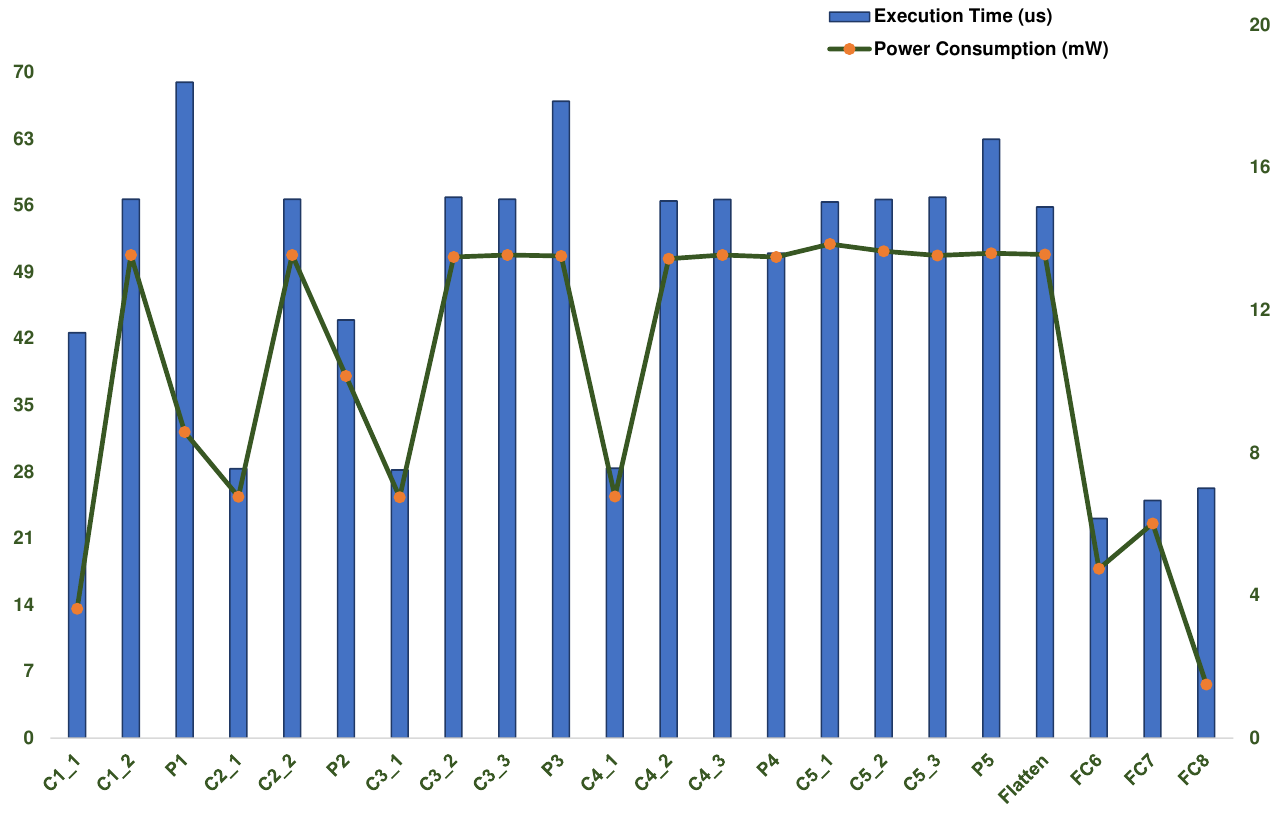}
    \caption{VGG-16 layer-wise execution time and power consumption.}
    \label{fig:vgg16_perf}
\end{figure}

For real-world validation, the vector engine is deployed on a Pynq-Z2 platform with an ARM Cortex-A9 host. The system achieves 154.6\,ms latency at 0.43\,W for object detection and classification workloads, outperforming prior works including LPRE~\cite{LPRE} (184\,ms / 0.93\,W), ILM~\cite{ILM} (163.7\,ms / 13.32\,W), EPSCON~\cite{GR-ACMTr} (772\,ms / 1.524\,W), and Flex-PE~\cite{Flex-PE} (186.4\,ms / 2.24\,W), as well as commercial platforms such as NVIDIA Jetson Nano (226\,ms / 1.34\,W) and Raspberry Pi (555\,ms / 2.7\,W). Fig.~\ref{fig:vgg16_perf} presents a layer-wise breakdown of execution time and power consumption for VGG-16, illustrating the benefits of precision-aware execution.

\section{Conclusion \& Future Work}

This work presented CARMEN, a runtime-adaptive vector engine that exploits configurable CORDIC iteration depth to dynamically trade off accuracy for efficiency in deep learning inference. The iterative MAC unit halves the FPGA resource footprint of prior CORDIC-based designs while consuming only 1.9\,mW, and the time-multiplexed multi-AF block supports seven nonlinear functions at less than 4\% area overhead. At the system level, a 256-PE ASIC configuration achieves 4.83\,TOPS/mm$^2$ compute density and 11.67\,TOPS/W energy efficiency in 28\,nm CMOS, while FPGA deployment on Pynq-Z2 delivers 154.6\,ms latency at 0.43\,W for real-time object detection, outperforming both prior academic accelerators and commercial edge platforms.

Future work will extend CARMEN into a compiler-assisted framework with automated layer-wise precision tuning and operation mapping, targeting end-to-end deployment of transformer-based workloads. Additionally, full physical design integration through place-and-route and exploration of multi-core vector engine configurations will be pursued to further improve scalability.


\bibliographystyle{ieeetr}
\bibliography{this_bibliography}

\end{document}